\documentclass{article}

\usepackage{proof}
\usepackage{amsmath}
\usepackage{amssymb}
\usepackage{stmaryrd}
\usepackage{natbib}

\newcommand{\splitc}{\,\check{\,}}
\newcommand{\bridge}{\,\hat{\,}}
\newcommand{\one}{\mathbf{1}}
\newcommand{\editout}[1]{}

\newcommand{\cslin}{\textit{CS}_{\textit{LIN}}}
\newcommand{\nlin}{\textit{NLIN}}

\newtheorem{theorem}{Theorem}[section]
\newtheorem{definition}[theorem]{Definition}

\newtheorem{lemma}[theorem]{Lemma}

\newtheorem{corollary}[theorem]{Corollary}

\newcommand{\qed}{\hfill $\Box$ \vspace{3mm}}

\usepackage{tikz}

\title{Type-logical grammars and formal languages}
\author{Richard Moot}

\begin{document}
\maketitle

\section{Introduction}

In a recent article, \citet[Section~4]{pshenitsyn2025first} provides a new lower bound for the string languages generated grammars in first-order linear logic, showing we can generate grammars in the class $\cslin${}, the context-sensitive languages generated in non-deterministic linear time.

The basic idea is modelling context-sensitive rewrites $AB \rightarrow BCD$ as Lambek calculus formulas $(A\bullet B)/(B\bullet C\bullet D)$, then using the Lambek calculus product to add a constant number $k$ of these formulas to each lexical entry of the grammar. So if $F$ is a first-order linear logic formula, and $A_1, \ldots A_j$ are  $j\leq k$ not necessarily different formulas of the form $(p_1\bullet\ldots \bullet p_m)/(q_1 \bullet \ldots\bullet q_n)$ (for all $p,q$ atomic), then we add $(\ldots (F \bullet A_1) \ldots \bullet A_j) $ for all choices of $j\leq k$ formulas $A$. 

This construction adds a number of rewrites to an existing grammar, each rewrite replacing a sequence of atomic formulas by another. Given that at most a constant number of rewrites is added to each word of the input sequence, at most a linear number of rewrites can be added in total. This gives a $\cslin${} lower bound on the string languages generated by grammars in first-order linear logic. We know $\textit{CFL} \subsetneq L(\cslin) \subsetneq \textit{CSL}$ \citep{book1978complexity}. Given that the context-sensitive languages are generated in non-deterministic linear \emph{space}, $L(\textit{MILL1}) \subseteq \textit{CSL}$, where equality would entail  $\textit{NP}=\textit{PSPACE}$. So the full picture of the string languages of first-order linear logic is as follows\footnote{The first inclusion is strict because the language $\{wcw^Rcw | w \in (a|b)^* \}$ is not in $\cslin$ but is a tree adjoining language, and therefore in MILL1. I conjecture that $L(\textit{MILL1}) \subsetneq \textit{CSL}$ as well, possibly with a condition that the first-order logic doesn't use term symbols.}.
\[
\textit{CFL} \subsetneq L(\cslin) \subsetneq L(\textit{MILL1}) \subseteq \textit{CSL}
\]

What makes the construction work is that in first-order linear logic, we can ensure that the rewrite rules commute freely. At its base, first-order linear logic is a commutative logic, and, where needed, non-commutativity is enforced by the use of string positions.

Given the result by \citet{book1978complexity} showing that fixed recognition for $\cslin${} is NP complete,   \citet{pshenitsyn2025first} showed that fixed recognition for MILL1 grammars is NP complete as well. MILL1 grammars therefore have NP complete \emph{fixed} recognition together with NP complete \emph{universal} recognition.

In this paper, we will do two things. First we will show that even though most type-logical grammars  operate wtih a non-commutative base logic, the mechanisms used for common linguistic constructions can be used to adapt the construction of \citet{pshenitsyn2025first} to any modern type-logical grammars which has product formulas, associativity\footnote{Or any form of modally licensed associativity.}, an extraction operator or any way to prefix constituents, and a non-deterministic infixation operation. We will show that any type-logical grammar formalism having all these ingredients will have NP complete fixed recognition as well.

Second, we will show how the quasi-linear languages $Q$, that is, the class of languages generated in non-deterministic linear time ($\textit{NSPACE}(n) = \nlin$, $Q = L(\nlin)$) can be generated by grammars in first-order linear logic. The quasi-linear languages are an interesting but powerful class of formal languages. The quasi-linear languages are the smallest abstract family of languages (AFL) closed under intersection. While \citet{pshenitsyn2025first} already showed that MILL1 grammars are closed under intersection, this shows that MILL1 grammars can also generate languages like the exponential language $a^{2^n}$, which can not be generated by intersections of context-free grammars.


This gives us new lower bounds for the class of languages generated by type-logical grammars, $L(\nlin)$ for first-order linear logic and $L(\cslin)$ for other type-logical grammars. 

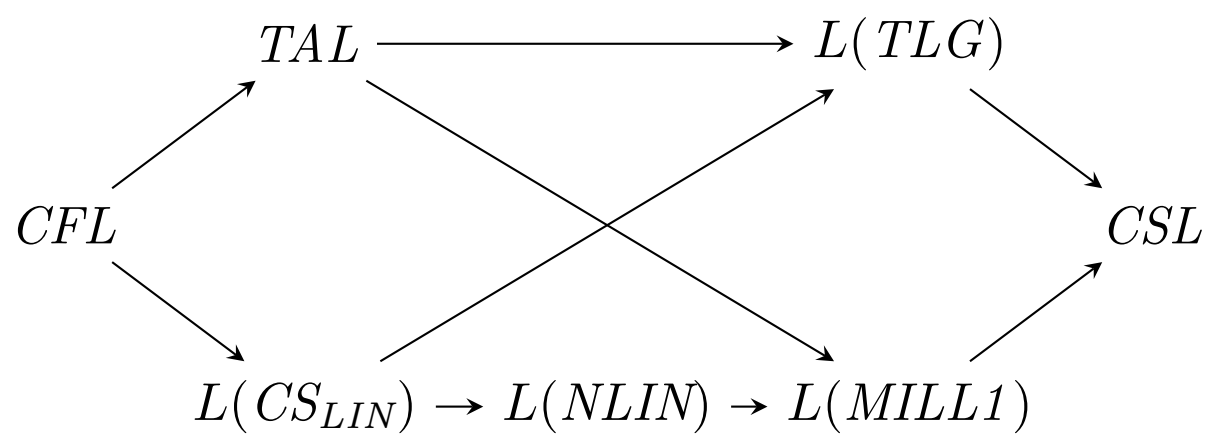
\begin{center}
\begin{tikzpicture}
\node (cfl) at (0,1.2) {$\mathit{CFL}$};	
\node (cslin) at (1.6,0) {$L(\cslin)$};
\node (tal) at (1.6,2.4) {$\mathit{TAL}$};	
\node (tlg) at (5.6,2.4) {$L(\mathit{TLG})$};	
\node (nlin) at (3.6,0) {$L(\nlin)$};	
\node (m1) at (5.6,0) {$L(\mathit{MILL1})$};
\node (csl) at (7.2,1.2) {$\mathit{CSL}$};
\draw [-stealth] (cfl) -- (cslin);
\draw [-stealth] (cslin) -- (nlin);
\draw [-stealth] (cslin) -- (tlg);
\draw [-stealth] (nlin) -- (m1);
\draw [-stealth] (m1) -- (csl);
\draw [-stealth] (cfl) -- (tal);
\draw [-stealth] (tal) -- (tlg);
\draw [-stealth] (tal) -- (m1);
\draw [-stealth] (tlg) -- (csl);
\end{tikzpicture}
\end{center}

It would seem that NLIN strictly contains TAL and probably also MCFL. We will see later that at least the iconic mildly context-sensitive languages are in NLIN. We will see that there are also a number of languages outside of these classes which can be generated by NLIN, notably the exponential language $a^{2^n}$ and languages whose relation with these classes is currently unknown, such as the generalised MIX languages.

NLIN should be contrasted to the better-known class LOGCFL. LOGCFL is the class of problems which are logspace many-one reducible to the context-free languages. In terms of automata, they can be seen as non-deterministic Turing machines with a single stack (like context-free grammars) and with a logspace auxiliary store \citep{sud78tape}. For NLIN, the auxiliary tapes have linear storage, but the time bound for computation is linear as well \citep{book1970}.

The precise relation between NLIN and LOGCFL is unknown and would seem to require important progress in complexity theory to be resolved, notably on the relation between logarithmic space with polynomial time and linear space with linear time.

\begin{center}
\begin{tabular}{|l|l|l|l|} \hline
Formalism & Storage & Time \\ \hline
CFL & 1 stack & polynomial \\
LOGCFL & 1 stack, 1 logspace Turing tape & polynomial 	\\
NLIN & 1 stack, 1 linear Turing tape & linear \\ \hline
\end{tabular}
\end{center}


For NLIN, we can see the three stacks as one stack (corresponding to the stack automata, like with context-free grammars) and one Turing tape (since two stacks can emulate one Turing tape). The key question is then whether one linear-space Turing tape with linear time is enough to emulate a log-space stack with polynomial time.

\section{Background}

In this section, we will briefly present the linear-time context-sensitive languages and the Lambek calculus. 

\subsection{The linear-time context-sensitive languages}

Although most work in formal language theory with applications to formal linguistics has focused on extensions of the context-free languages, some authors, notably \citet{book1978complexity} and \citet{dw86growing} have looked at restrictions of the context-sensitive languages instead.

There are a number of  equivalent ways of defining context-sensitive grammars. However, since we are interested in linear-time context-sensitive grammars, we choose a very general definition.

\begin{definition}
A context-sensitive grammar is a phrase structure grammars where all rules are of one of the following two forms
\begin{align*}
   \alpha\rightarrow \beta \\
	A \rightarrow a
\end{align*}	
Where $\alpha$ and $\beta$ are non-empty sequences of non-terminals such that $|\alpha| \leq |\beta|$ 
\end{definition}

It is well-known that the context-sensitive languages correspond to linear-bounded Turing machines, that is, the context-sensitive languages can be recognised in $\textit{NSPACE}(n)$.

 \citet{book1978complexity} studies the languages generated by context-sensitive grammars in  $\textit{NTIME}(n)$, a class he calls $\cslin$. He shows that $\cslin${} is strictly included in  $\textit{NTIME}(n)$ and that fixed recognition for $\cslin${} is NP-complete.
 
\subsection{The Lambek calculus}

The Lambek calculus was introduced by \citet{lambek58} as a logic designed for parsing natural language sentences. It is based on the idea that a grammatical sentence corresponds to a theorem in the Lambek calculus, through the intermediary of a \emph{lexicon} which is a function from words to sets of formulas.

The Lambek calculus starts from a fixed set of atomic formulas. In a linguistic context, these often correspond to basic grammatical categories: $n$ for common noun, $np$ for noun phrase and $s$ for sentence.  We can then form complex formulas using three connectives.

The connective  $A/B$ looks for a $B$ to its right to form an $A$. So a formula $np/n$ looks for a noun to its right to form  a noun phrase. 

Similarly, the connective $B\backslash A$ looks for a $B$ to its left to form an $A$. For example, a formula $np\backslash s$ looks for a noun phrase to its right to form a sentence.

Finally, the connective $A\bullet B$ denotes the concatenation of $A$ with $B$. While it is a part of the logic, it is rarely used for grammars.

Figure~\ref{fig:lambek} shows the rules of the (cut-free) sequent calculus for the Lambek calculus.

Now, given a lexicon. 
\begin{align*}
\textit{lex}(\textrm{a}) &= \{ np/n \} \\	
\textit{lex}(\textit{program}) &= \{ n \} \\
\textit{lex}(\textit{halts}) &= \{ np\backslash s\}
\end{align*}

\begin{figure}
\begin{center}
\begin{tabular}{ccc}
&\infer[\textit{Ax}]{A\vdash A}{} \\[2mm]
Irreverisble,  asynchronous && Reversible, synchronous \\[2mm]
\infer[\backslash L]{\Gamma,\Delta,B\backslash A,\Gamma'\vdash C}{\Delta\vdash B & \Gamma,A,\Gamma'\vdash C} &&
\infer[\backslash R]{\Gamma\vdash B\backslash A}{B,\Gamma\vdash A}\\[2mm]
\infer[/ L]{\Gamma,A/B,\Delta,\Gamma'\vdash C}{\Gamma,A,\Gamma'\vdash C & \Delta\vdash B} && 
\infer[/ R]{\Gamma\vdash A/B}{\Gamma,B\vdash A} \\[2mm]
\infer[\bullet R]{\Gamma,\Delta\vdash A\bullet B}{\Gamma\vdash A & \Delta\vdash B} &&
\infer[\bullet L]{\Gamma,A\bullet B,\Gamma'\vdash C}{\Gamma,A,B,\Gamma'\vdash C}
\end{tabular}
\end{center}
\caption{Sequent calculus rules for the Lambek calculus}
\label{fig:lambek}	
\end{figure}

The following proof shows that, given the lexicon above,  the sentence `a program halts' is grammatical.
\[
\infer[\backslash L]{np/n, n,  np\backslash s\vdash s}{\infer[\textit{Ax}]{s\vdash s}{} & 
\infer[/ L]{np/n, n\vdash np}{
\infer[\textit{Ax}]{np\vdash np}{} & 
\infer[\textit{Ax}]{n\vdash n}{} 
}
}
\]

\citet{pentus} proved that Lambek grammars generate exactly the context-free languages, but even before this many  authors worked on extensions of the Lambek calculus more suitable for natural language while preserving its good logical properties, such as cut elimination and model theory. I will use the general name \emph{type-logical grammars} for this research program.

Focusing proof search was introduced for linear logic as a way of eliminating the inessential rule permutations from sequent calculus proof search \citep{focusb}.

\begin{definition} A proof is \emph{focused} if the proofs alternates between reversible and irreversible phases until it reaches the atoms as follows. 
\begin{enumerate}
\item\label{step:1} If our sequent is of the form $A\vdash A$ with $A$ an atomic formula, this subproof is correct.
	\item If a reversible rule can be applied to the sequent, we apply this rule.
	\item If no reversible rules can be applied, select a formula and focus it. We apply irreversible rules on this focused formula and its subformulas until we reach either atoms or reversible subformulas, continuing the proof with step~\ref{step:1}.
\end{enumerate}
	
\end{definition}

\begin{lemma} If $\Gamma\vdash C$ is derivable, then it is derivable with a focused proof.
\end{lemma}

\section{Type-logical grammars and linear-time context sensitive languages}

In this section we show that all type-logical grammars with the following elements have an NP-complete fixed recognition problem.
\begin{enumerate}
    \item\label{it:ass} Associativity, including partial associativity, such as modally licensed versions of associativity
	\item\label{it:prod} The Lambek calculus product rules $\bullet L$ and $\bullet R$
	\item\label{it:extr} Medial extraction
	\item\label{it:infx} Non-deterministic infixation
\end{enumerate}

Items~\ref{it:ass} and~\ref{it:prod}, the product rules and associativity are the standard rules for the Lambek calculus. In what follows, we will first show that a logic having all four of the above elements has an NP-complete fixed recognition problem, using the construction of \citet{pshenitsyn2025first}. Then we will show that many prominent type-logical grammars satisfy all these four conditions. The first three are relatively standard, given that most of the extensions of the Lambek calculus have been motivated by giving a better account of medial extraction than the Lambek calculus. While an explicit non-deterministic infixation rule is not as common, we will see that it is relatively easy to implement this behaviour given the standard tools of these logics.

\subsection{Extraction and infixation}

\citet{moortgat88phd,m96q} proposes the following rules for extraction and infixation\footnote{Moortgat uses $A\downarrow B$ where we use $B\Downarrow A$ (note the swapped argument and result), and $A\uparrow B$ where we use $A\Uparrow B$. We changed the notation both to distinguish these connectives from the ones in the Displacement calculus and to highlight the similarities.} in extensions of the Lambek calculus. 

\begin{align}
\infer[\Downarrow L]{\Gamma,\Delta, B\Downarrow A,\Delta',\Gamma'\vdash C}{\Delta,\Delta' \vdash B & \Gamma, A, \Gamma' \vdash C}
&&
\infer[\Uparrow R]{\Gamma, \Gamma' \vdash C\Uparrow A}{\Gamma, A, \Gamma' \vdash C}
\end{align}

He notes that the combination $(s\Uparrow np) \Downarrow s$ almost behaves like a generalised quantifier, modelling Montague's `quantifying in' operation but in the absence of a mechanism to ensure the noun phrase and the quantifier occur in the same space. In other words, while the following pattern is valid.
\begin{align}\label{eq:lr}
\infer[\Downarrow L]{\Gamma,\Delta, (s\Uparrow np)\Downarrow s,\Delta',\Gamma'\vdash C}{\infer[\Uparrow R]{\Delta,\Delta' \vdash s\Uparrow np}{\Delta, np,\Delta' \vdash s} & \Gamma, A, \Gamma' \vdash C}
\end{align}
However, the $\Uparrow R$ rule allows us to insert the $np$ anywhere, not just at the position of the original $(s\Uparrow np) \Downarrow s$ formula.

Many solutions have been proposed to turn a system with rules like $\Uparrow R$ and $\Downarrow L$ into a full logic, adding both rules which would correspond to $\Uparrow L$ and $\Downarrow R$ and a mechanism to ensure identity between the extraction and infixation point in the rule combination shown in~\ref{eq:lr}. We will see a few of those in what follows. The important point is that these extensions still allow us to emulate the $\Uparrow R$ and $\Downarrow L$ rules and that the non-determinism of the $\Downarrow L$ insertion is one of the key elements which allows us to simulate string rewrites.
 
\subsection{String rewrites with extraction and infixation}
To simulate the analysis of \citet{pshenitsyn2025first}, we use a two-step process. Lexical formulas are assigned formulas of the form $G_1 \bullet \ldots \bullet G_n \bullet p$ where $p$ is an atomic formula, and all $G$ are formulas which move in two steps. First, the $G_i$ of all lexical entries move a subformula $F_i$ to the front of the sequent, leaving us with a sequent of the following form
\[
F_1, \ldots, F_n, p_1, \ldots, p_m \vdash s
\]
Then, each $F_i$ formula inserts a subformula $(q_1\bullet \ldots\bullet q_i)/(r_1\bullet \ldots \bullet r_j)$ between the $p_i$ allowing it to apply the rewrite $r_1,\ldots r_j \rightarrow q_1,\ldots q_i$. We will write $\delta/\gamma$ for a formula representing one of this rewrites.

%

For the first step, we need to show that we can move the $F_i$ to the front of the sequent while keeping $p$ in place. This is done as follows.
\begin{align}\label{eq:prefixf}
\infer[\bullet L]{\Gamma,((F\backslash s)\Downarrow s)\bullet p,\Gamma'\vdash s}{
\infer[\Downarrow L]{\Gamma,(F\backslash s)\Downarrow s, p,\Gamma'\vdash s}{
 \infer[\backslash R]{\Gamma, p, \Gamma' \vdash F\backslash s}{
   \infer*{F,\Gamma, p, \Gamma' \vdash s}{}
 }
&
\infer[\textit{Ax}]{s\vdash s}{}
}
}
\end{align}

For the second step, we take one of the possible rules and move it to the left of a sequence of atomic formulas which corresponds to the left-hand side of its rule.
\begin{align}\label{eq:infixanywhere}
\infer[\Downarrow L]{\Gamma, (s\Uparrow (\delta/\gamma))\Downarrow s, \Gamma',  \Delta, \Delta' \vdash s}{
\infer[\Uparrow R]{\Gamma, \Gamma', \Delta, \Delta' \vdash s\Uparrow (\delta/\gamma)}{
\infer[/ L]{\Gamma, \Gamma', \delta/\gamma, \Delta, \Delta' \vdash s}{
\infer*{\Delta \vdash \gamma}{}
&
\infer*{\Gamma, \Gamma', \delta, \Delta' \vdash s}{}
}
}
&
\infer[\textit{Ax}]{s\vdash s}{}
}
\end{align}

If we look at the $/L$ rule in the top left of~\ref{eq:infixanywhere} in more detail, it looks as follows (where $\Delta$ is $r_1,\ldots,r_j$). 
\begin{align}
\infer[/L]{\Gamma,\Gamma', (q_1\bullet \ldots\bullet q_i)/(r_1\bullet \ldots \bullet r_j), r_1,\ldots,r_j,\Delta'\vdash s}{\infer[\bullet R\ \text{($j$ times)}]{r_1,\ldots,r_j\vdash r_1\bullet \ldots \bullet r_j }{\infer[\textit{Ax}]{r_1\vdash r_1}{}\;\ldots\; \infer[\textit{Ax}]{r_j\vdash r_j}{}} & \infer[\bullet L \text{($i$ times)}]{\Gamma,\Gamma',q_1\bullet \ldots\bullet q_i,\Delta'\vdash s}{\infer*{\Gamma,\Gamma',q_1,\ldots,q_i,\Delta'\vdash s}{}}}
\end{align}

The net result is that we have eliminated one of the rules, and performed one rewrite step.

%

A lexical entry for a word assigned atomic formula $p_k$ in the lexicon then has the general form
\begin{align}\label{eq:lex}
((F_1\backslash s)\Downarrow s) \bullet \ldots \bullet ((F_n\backslash s)\Downarrow s) \bullet p_k
\end{align}
\noindent with each $F_i$ of the following form
\begin{align}
(s\Uparrow (\delta_i/\gamma_i))\Downarrow s	
\end{align}
with $\delta_i/\gamma_i$ representing a rule instance from $R$.

Now it is easy to establish the following invariant: the atomic $p$ formulas represent the current sequence of non-terminal symbols, whereas all $\delta/\gamma$ subformulas represent the context-sensitive rules still to be applied.

\begin{lemma}
If a sequent containing entries of the form shown in~\ref{eq:lex} is derivable with goal formula $s$, then it has a proof which proceeds in the follow three phases.
\begin{enumerate}
\item $\bullet L$ phase, where separate the $p$ atomic formulas from the prefixation formulas $(F\backslash s)\Downarrow s$.
\item\label{it:phase:prefix} A prefixation phase, moving the $F_i$ formulas to the front of the sequent using the $\Downarrow L$ and $\backslash R$ rules as shown in~\ref{eq:prefixf}.
\item\label{it:phase:infix} An infixation phrase, moving the $\delta/\gamma$ subformulas as shown in~\ref{eq:infixanywhere}.
\end{enumerate}
\end{lemma}

\paragraph{Proof} Focusing proof search guarantees we can apply all $\bullet L$ rules first, but also that both the combinations of $\Downarrow L$ and and $\backslash R$ of phase~\ref{it:phase:prefix} and the combination of $\Downarrow L$, $\Uparrow R$, $\bullet L$ and $\bullet R$ of phase~\ref{it:phase:infix} are a single alternation of irreversible and reversible rules. So we only have to show that any infixations which occur before a prefixation can be permuted. We proceed by induction on the number of infixations which are followed by prefixations. If there are none, we are done. If not, we look at the topmost infixation and the lowest prefixation. This gives the following situation schematically
\[
\infer[\Downarrow L]{\Gamma, (s\Uparrow (\delta/\gamma))\Downarrow s, \Gamma',  \Delta, \Delta', (F\backslash s)\Downarrow s, \Delta'' \vdash s}{
\infer[\Uparrow R]{\Gamma, \Gamma', \Delta, \Delta', (F\backslash s)\Downarrow s, \Delta'' \vdash s\Uparrow (\delta/\gamma)}{
\infer[/ L]{\Gamma, \Gamma', \delta/\gamma, \Delta, \Delta', (F\backslash s)\Downarrow s, \Delta'' \vdash s}{
\infer*{\Delta \vdash \gamma}{}
&
\infer[\Downarrow L]{\Gamma, \Gamma', \delta, \Delta', (F\backslash s)\Downarrow s, \Delta'' \vdash s}{\infer[\backslash R]{\Gamma, \Gamma', \delta, \Delta', \Delta'' \vdash F\backslash s}{\infer*{F,\Gamma, \Gamma', \delta, \Delta', \Delta'' \vdash s}{}} & \infer[\textit{As}]{s\vdash s}{}}
}
}
&
\infer[\textit{Ax}]{s\vdash s}{}
}
\]
The top two rules $\backslash R$ and $\Downarrow L$ are the prefixation part and we can move them to be the bottom two rules of the subproof as follows.
\[
\infer[\Downarrow L]{\Gamma, (s\Uparrow (\delta/\gamma))\Downarrow s, \Gamma',  \Delta, \Delta', (F\backslash s)\Downarrow s, \Delta'' \vdash s}{
\infer[\backslash R]{\Gamma, (s\Uparrow (\delta/\gamma))\Downarrow s, \Gamma',  \Delta, \Delta', \Delta'' \vdash F\backslash s}{
 \infer[\Downarrow L]{F,\Gamma, (s\Uparrow (\delta/\gamma))\Downarrow s, \Gamma',  \Delta, \Delta', \Delta'' \vdash s}{
 \infer[\Uparrow R]{F,\Gamma, \Gamma', \Delta, \Delta', \Delta'' \vdash s\Uparrow (\delta/\gamma)}{
   \infer[\backslash L]{F,\Gamma, \Gamma', \delta/\gamma, \Delta, \Delta', \Delta'' \vdash s}{
   \infer*{\Delta \vdash \gamma}{}
   & \infer*{F,\Gamma, \Gamma', \delta, \Delta', \Delta'' \vdash s}{}
   }
 }
 &
\infer[\textit{Ax}]{s\vdash s}{}
 }
}
&
\infer[\textit{Ax}]{s\vdash s}{}
}
\]
The case where the $\delta/\gamma$ formula is infixed into $\Gamma'$ is symmetric. \qed

\subsection{The Displacement calculus}

The Displacement calculus of \citet{mvf11displacement} provide one way to turn the extraction and infixation connectives of \citet{m96q} into a full logic. This is implemented by using a `marker' $\one$ to indicate the extraction point used by the extraction rule. The infixation rule then replaces this marker by the formula to be infixed. This combination of rules is then used for the treatment of generalised quantifiers and other linguistic phenomena.
\begin{align}
\infer[\uparrow R]{\Gamma, \one, \Gamma' \vdash C\uparrow A}{\Gamma, A, \Gamma' \vdash C}
&&
\infer[\downarrow L]{\Gamma,\Delta, B\downarrow A,\Delta',\Gamma'\vdash C}{\Delta, \one,\Delta' \vdash B & \Gamma, A, \Gamma' \vdash C}
\end{align}
However, the Displacement calculus also provides ways both to remove this extraction symbol (necessary for the treatment of extraction) and to add one. 
\begin{align}
\infer[\splitc R]{\Gamma,\one,\Gamma \vdash \bridge C}{\Gamma, \Gamma' \vdash C}	&&
\infer[\bridge R]{\Gamma,\Gamma'\vdash \splitc C}{\Gamma, \one, \Gamma' \vdash C}
\end{align}

We can simulate $C\Uparrow A$ as $\bridge(C\uparrow A)$ and $B\Downarrow A$ as $\splitc B \downarrow A$ as follows.
\begin{align}
\infer[\bridge R]{\Gamma, \Gamma' \vdash \bridge(C\uparrow A)}{\infer[\uparrow R]{\Gamma, \one, \Gamma' \vdash C\uparrow A}{\Gamma, A, \Gamma' \vdash C}}
&&
\infer[\downarrow L]{\Gamma,\Delta, \splitc B\downarrow A,\Delta',\Gamma'\vdash C}{\infer[\splitc R]{\Delta, \one,\Delta' \vdash \splitc B}{\Delta, \Delta' \vdash B} & \Gamma, A, \Gamma' \vdash C}	
\end{align}
Both are standard idioms in the Displacement calculus: $\bridge(C\uparrow A)$ is used in the analysis of extraction, and $\splitc B \downarrow A$  for parentheticals \cite[Sections~3.2.2--3.2.5]{mvf11displacement}.

A lexical entry for a word assigned atomic formula $p_k$ in the lexicon then has the general form
\begin{align}
(\splitc(F_1\backslash s)\downarrow s) \bullet \ldots \bullet (\splitc(F_n\backslash s)\downarrow s) \bullet p_k
\end{align}
Each $F_i$ is then of the following form.
\begin{align}
\splitc(\bridge(s\Uparrow (\delta_i/\gamma_i)))\Downarrow s	
\end{align}
With $\delta_i/\gamma_i$ representing a rule instance from $R$.

\subsection{The associative logic of scope}

\citet{bs14cont} and \citet{barker2019nllam} propose the logic the logic of scope $\textit{NL}_{\lambda}$. While this logic is non-associative, we are interested here in the version of the logic associativity added, $\textit{L}_{\lambda}$. This translation does not extend to the original, non-associative logic since the formulas $\delta_i/\gamma_i$ rewrite trees of atoms instead of sequences on atoms as required for 	the translation to work.  

The set of rules needed for what follows is listed below. It is a relatively standard multimodal categorial logic with two families of connectives, and a connective $t$ representing the empty string. What distinguishes the logic from others are the $\beta$ and $\beta^{-1}$ rules. Inspired by $\beta$ reduction and expansion for the lambda calculus, these rules allow use to move a constituent $\Delta$ into and out of a position marked by lambda abstraction.

\begin{align*}
\infer[\beta]{\Xi[\Delta \mathbin{\circledcirc}\lambda x.\Gamma[x]]\vdash C}{\Xi[\Gamma[\Delta]]\vdash C} &&
\infer[\beta^{-1}]{\Xi[\Gamma[\Delta]]\vdash C}{\Xi[\Delta \mathbin{\circledcirc}\lambda x.\Gamma[x]]\vdash C}\\
\\
\infer[\epsilon]{\Gamma,\epsilon,\Delta\vdash C}{\Gamma,\Delta\vdash C} &&
\infer[\epsilon]{\Gamma,\Delta\vdash C}{\Gamma,\epsilon,\Delta\vdash C}
\\
\\
\infer[t L]{\Gamma, t, \Delta\vdash C}{\Gamma,\epsilon,\Delta\vdash C} &&
\infer[t R]{\epsilon\vdash t}{} \\
\infer[\fatslash L]{\Gamma[C\fatslash A \circledcirc \Delta]\vdash D}{\Delta\vdash A & \Gamma[C]\vdash D} &&
\infer[\fatbslash R]{\Gamma\vdash A\fatbslash C}{A\circledcirc \Gamma\vdash C} &&
\\
\infer[\varodot R]{\Gamma\circledcirc\Delta\vdash A\varodot B}{\Gamma\vdash A & \Delta\vdash B} 
\infer{}{}
\end{align*}

For the associative version of Barker's logic of scope, we can define $C\Uparrow A$ as $t \varodot (A\fatbslash C)$. As in the Displacement calculus, this is just the standard treatment of extraction in the logic.

\begin{align}
\infer[\epsilon]{\Gamma, \Gamma'\vdash t \varodot (A\fatbslash C)}{
\infer[\beta]{\Gamma, \epsilon, \Gamma'\vdash t \varodot (A\fatbslash C)}{
\infer[\varodot R]{\epsilon \mathbin{\circledcirc} \lambda x. (\Gamma, x, \Gamma')\vdash t \varodot (A\fatbslash C)}{
\infer[tR]{1\vdash t}{} &
\infer[\fatbslash R]{\lambda x. (\Gamma, x, \Gamma')\vdash  A\fatbslash C}{
\infer[\beta^{-1}]{A \circledcirc \lambda x. (\Gamma, x, \Gamma')\vdash  C}{
\Gamma,  A, \Gamma'\vdash C
}
}
} 	
}
}
\end{align}

Similarly, we can define $B\Downarrow A$ as $A\fatslash (t\fatbslash B)$

\begin{align}
\infer[\beta]{\Gamma, \Delta, A\fatslash (t\fatbslash B),\Delta',\Gamma'\vdash C}{
\infer[\fatslash L]{\Gamma, (A\fatslash (t\fatbslash B)) \circledcirc \lambda x. (\Delta, x, \Delta'),\Gamma'\vdash C}{
\infer[\fatbslash R]{\lambda x. (\Delta, x, \Delta')\vdash t\fatbslash B}{
\infer[t L]{t \circledcirc \lambda x. (\Delta, x, \Delta')\vdash B}{
\infer[\beta^{-1}]{\epsilon \circledcirc \lambda x. (\Delta, x, \Delta')\vdash B}{
\infer[\epsilon]{\Delta, \epsilon,\Delta' \vdash C}{
\Delta, \Delta' \vdash C
}
}
}
}
& \Gamma, A, \Gamma' \vdash C
}
}	
\end{align}


A lexical entry for a word assigned atomic formula $p_k$ in the lexicon then has the general form
\begin{align}
(s\fatslash (t\fatbslash(F_1\backslash s))) \bullet \ldots \bullet (s\fatslash (t\fatbslash(F_n\backslash s))) \bullet p_k
\end{align}
Each $F_i$ is then of the following form.
\begin{align}
s\fatslash (t\fatbslash (t\varodot ((\delta_i/\gamma_i)\fatbslash  s)))
\end{align}
With $\delta_i/\gamma_i$ representing a rule instance from $R$.


\subsection{Hybrid type-logical grammars}

Hybrid type-logical grammars were introduced by \cite{kl12gap,kl20tls} as a way of combining the standard Lambek calculus connectives with lambda terms for more flexible string operations \citep{oehrle}.

Because of the way lambda calculus connectives connect with the other connectives of the logic, hybrid type-logical grammars do not allow formulas of the form $(A\multimap B)\bullet C$, since the lambda calculus connective `$\bullet$' cannot have a subformula containing the linear connective `$\multimap$'. However, we can use lexical entries which would correspond to $((F\backslash s)\uparrow p)\downarrow s$ in the Displacement calculus. This will give us formulas of the following schematic form $(((F_1\backslash \ldots (F_n\backslash s))\uparrow p)\downarrow s$ 

Lexical entries of this form use standard quantifying in for the $p$ formula, ensuring the $p$ formula will return to the place of the full formula, with the $\backslash$ connective ensuring  that the $F_i$ formulas appear at the front.

This alternative analysis looks as follows in the Displacement calculus.
\begin{align}\label{eq:altdisp}
\infer[\downarrow L]{\Gamma,((F\backslash s)\uparrow p)\downarrow s,\Gamma'\vdash s}{
\infer[\uparrow R]{\Gamma, \one, \Gamma' \vdash (F\backslash s)\uparrow p}{
 \infer[\backslash R]{\Gamma, p, \Gamma' \vdash F\backslash s}{
   \infer*{F,\Gamma, p, \Gamma' \vdash s}{}
 }
}
&
\infer[\textit{Ax}]{s\vdash s}{}
}
\end{align}
Compare proof~\ref{eq:altdisp} to proof~\ref{eq:prefixf} to verify we obtain the same conclusion from the same premisses in both proofs.

Translating this analysis to hybrid type-logical grammars is done as follows. As a first step, the formula above is translated as follows.
\[
(p \multimap F\multimap s) \multimap s
\] 
Then, $F$ will be of the following form.
\[
((\delta/\gamma) \multimap s)\multimap s
\]
The missing part is then the lambda term to assign to this formula. For a word $w$ in the lexicon, the following term does everything we need.
\[
\lambda P.  ((P\ w) (\lambda Q. (Q\, \epsilon))) :  (p \multimap ( ((\delta/\gamma) \multimap s)\multimap s ) \multimap s) \multimap s
\]
The first argument of $P$ is the word $w$ in a standard quantifying-in analysis, then the second argument of $P$ simply ensures that the string corresponding to the formula $\delta/\gamma$ is the empty string.  So this term, while looking a bit complicated, is simply the combination of a standard treatment of quantifying-in with a standard treatment of extraction.

The following proof and beta reduction sequence show that everything works as intended. 
\newcommand{\sphantom}{\phantom{(}s\phantom{)}}
\newcommand{\fform}{((R\multimap s)\multimap s)}
\[
\infer[\multimap E]{\sphantom}{\infer[\multimap I^2	]{p\multimap (\fform\multimap s)}{
\infer[\multimap I^3]{\fform\multimap s}{
\infer[\multimap E]{\sphantom}{
\infer[\multimap I^1]{R\multimap s}{\infer*{\sphantom}{\Delta, [R]^1, \Gamma, [p]^2, \Gamma'}
} & [(R\multimap s)\multimap s]^3}}
} & (p \multimap (\fform\multimap s))\multimap s}
\]

\begin{align*}
(\lambda P.  ((P\ w) (\lambda Q. (Q\, \epsilon))))\, (\lambda z. \lambda x. (x\, (\lambda y. M[y,z]))) \\
 (((\lambda z. \lambda x. (x\, (\lambda y. M[y,z]))) w) (\lambda Q. (Q\, \epsilon)))) \\
 (\lambda x. (x\, (\lambda y. M[y,w]))) (\lambda Q. (Q\, \epsilon))\\
 \lambda Q. (Q\, \epsilon)) (\lambda y. M[y,w]) \\
 (\lambda y. M[y,w])\, \epsilon\\
 M[\epsilon,w]
\end{align*}

In full generality, we need multiple formula $F_i$ and therefore the following formula.
\[
(p \multimap F_1 \ldots \multimap  F_n \multimap s) \multimap s
\] 
\noindent with each $F_i$ as before.
\[
((\delta/\gamma)\multimap s)\multimap s
\]
The corresponding lambda term then looks as follows, applying the standard extraction pattern to each of the $Q_i$ corresponding to $F_i$ and the quantifying in pattern to $p$.
\begin{align*}
\lambda P.  ((P\ w) (\lambda Q_1. (Q_1\, \epsilon)) \ldots (\lambda Q_n. (Q_n\, \epsilon))) : \\ (p \multimap  (((\delta_1/\gamma_1) \multimap s)\multimap s ) \ldots \multimap (((\delta_n/\gamma_n)\multimap s)\multimap s)   \multimap s) \multimap s
\end{align*}

\subsection{Multimodal type-logical grammars}

Finally, for multimodal type-logical grammars, we can use a much simpler solution. In a multimodal logic, we can use the idiom $\Diamond\Box A$ to allow a formula to permute throughout the sequent, then function as an $A$ formula after its movement. 

The key logical and structural rules making this possible are listed below.
\begin{align*}
\infer[\Diamond L]{\Gamma, \Diamond A,\Gamma' \vdash C}{\Gamma, \langle A\rangle, \Gamma'\vdash C} &&
\infer[\Box L]{\Gamma, \langle \Box A\rangle, \Gamma'\vdash C}{\Gamma, A,\Gamma'\vdash C} \\
\infer[\textit{C}]{\Gamma,\langle \Delta\rangle, \Delta', \Gamma'\vdash C}{\Gamma,\Delta',\langle \Delta\rangle,\Gamma' \vdash C}
&&
\infer[\textit{C}]{\Gamma,\Delta',\langle \Delta\rangle,\Gamma' \vdash C}{\Gamma,\langle \Delta\rangle, \Delta', \Gamma'\vdash C}
\end{align*}
This assumes an associative base logic. However, we can adapt to the non-associative case by simple adding a second pair of unary modalities for controlled associativity, as shown by \citet{KurtoMM}.

 We then assign formulas of the following form to the lexical entries. 
 \[(
\ldots (p \bullet \Diamond\Box (\delta_1/\gamma_1)) \ldots \Diamond\Box (\delta_k/\gamma_k)
\]
This analysis is simpler than for previously discussed systems (and close to the way \citet{pshenitsyn2025first} implements this for first-order linear logic) because we have no need of forms lifting to permute subformulas, and we can use the permute modality both to move the formulas corresponding to context-sensitive rules `out of the way' (so they don't block the application of another context-sensitive rule) and to move them to the correct position when their rule needs to be applied.

Compared to the results of 	\citet{diss}, who gives a tight bound of CSL for multimodal type-logical grammars, we use a more standard and more restricted set of structural rules.

\subsection{Corollaries}

\citet{pshenitsyn2025first} used Theorem~1 from \citet{book1978complexity}, showing that there are fixed languages in $\cslin$ for which the decision problem is NP complete, to show that fixed recognition for string grammars in first-order linear logic is NP complete as well.
As a consequence of the translation of $\cslin$ grammars into the Displacement calculus, the associative logic of scope, and hybrid type-logical grammars, fixed recognition for all these formalisms is NP-complete as well. 

\begin{corollary}
	Fixed recognition for type-logical grammars in NP complete
\end{corollary}

While 

\begin{corollary}
	Unless P=NP, type-logical grammars are more powerful than mild\-ly context-sensitive formalisms.
\end{corollary}

One of the defining characteristics of mildly context-sensitive formalisms is polynomial fixed recognition\footnote{While some authors prefer polynomial parsing, I prefer polynomial fixed recognition, since it is more clearly a property of classes of formal languages.}

For multimodal type-logical grammars,  this gives a $\cslin$ lower bound and a CSL upper bound, plus an NP-hard fixed recognition problem. Finding a tight upper bound for the other formalisms remains an important open question.

All these formalisms are therefore quite powerful, and it makes sense to look for restrictions ensuring polynomial or LOGCFL fixed recognition.

The following language is not in $\cslin$ \citep{book1978complexity}.
\[
L = \{ w c w^R c w \;|\; w \in (a,b)^* \}
\]
In this language, $w$ ranges over sequences of $a,b$, and this sequence is both reversed (indicated by $w^R$) and copied. However, this language is a simple tree adjoining language, and can therefore be generated by the type-logical grammars we've discussed in this paper\footnote{This language can be generated in \emph{deterministic} linear time (LIN): using three pointers $p$, $q$, $r$, start with $p, q, r$ at the beginning of the string, move $q$ and $r$ right until the second $c$, then move $p$ and $r$ forward, $q$ backward and verify all three pointers point to the same symbol, succeed if $p$ and $q$ are at the same position pointing to $c$ and $r$ is at the end of the string.  Fail in all other cases.}.	

\subsection{Possible restrictions}

Summarising the results so far,  we have shown that the results of \citet{pshenitsyn2025first} can be adapted to many other type-logical grammars, showing these formalisms generate a larger class of string languages than was previously known.  

The construction pushes the limits of these formalisms by exploding the size of the lexicon.

\begin{enumerate}
	\item The $\bullet L$ and $\bullet R$ are needed for modelling the rewrite rules beyond context-free: $\bullet R$ is needed for complex right-hand sides of rules, whereas $\bullet L$ is needed to decompose a previously generated formula into a sequence of atoms so that a next rule can apply.
	\item While medial extraction would be difficult to eliminate because of the many linguistic applications, I believe this is less clear for the arbitrary infixation operation exemplified by Moortgat's $\Downarrow L$ and Morrill e.a.'s $\splitc$, where applications are more limited.   
	 However,
	$\splitc$ is definable as $A\uparrow I$, and we cannot just remove the connective $I$ from the Displacement calculus, or $t$ from the logic of scope, since these connectives are used for extraction as well. 
	\item Associativity is needed to model the string rewrites. Switching to a fully non-associative system is possible, but this means losing some of the benefits of associativity for treating linguistic phenomena like right-node raising. Absent associativity, the current construction would amount to a linear-time tree rewrite system, and that still seems powerful. 
	\item We could restrict the \emph{order} of formulas in the lexicon. The construction as given in this paper uses formulas with order 5. This is a bit high\footnote{If we calculate the order of formulas containing products in the same way \citet{pentus10poly} does, the order of our formulas becomes 6, since the order of a formula $(p\bullet q)/r$ is 2 instead of 1. I like this way of calculating order, since it ensures formulas of order 1 are essentially Horn clauses.}. In general, order 3 or 4 suffices for grammatical purposes. So unless the construction can be made to work with lower-order formulas, this gives a possible restriction. While the order restriction is reasonable for hybrid type-logical grammars, the Displacement calculus and the associative logic of scope, for multimodal type-logical grammars and first-order linear logic, the construction uses formulas of very low order: order 1 according to the standard calculation, and order 2 when using Pentus's additional term for products. 
	\item Best-case scenario would be a restriction to \emph{growing} context-sensitive grammars \citep{dw86growing}. This would guarantee polynomial parsing for fixed grammars and LOGCFL complexity. While this is easy to enforce on formulas of the form $(p_1 \bullet \ldots \bullet p_i)/(q_1 \bullet \ldots \bullet q_j)$ by requiring $j > i$, this restriction would need to be generalised is some way to all formulas in our logic. However, I doubt that there are reasonable restrictions on our logics which would ensure the planarity requirement of proof traces needed for the complexity result of \citet{dw86growing}. That is, restrictions which would enforce the planarity requirement without giving up the expressivity needed for linguistic applications.
	\item Another best-case scenario would be an extension of the results of \citet{pnlinear} and \citet{murong} to show linear time verification of correctness for proof nets in first-order linear logic. This would show that the class of languages generated in first-order linear logic is exactly characterised by the class NLIN.
\end{enumerate}

\section{NLIN grammars in first-order linear logic}

\subsection{NLIN automata and their grammars}

\citet{book1970} study the class of formal languages generated by non-de\-ter\-mi\-nistic Turing machines in linear time, i.e.\ $\textit{NSPACE}(n)$. They call this class \nlin, and show that this complexity class corresponds to a number of different classes of automata. They call the class of languages generated by such automata $Q$, the \emph{quasi-linear languages}.

There are multiple equivalent ways to define NLIN automata, using stacks and/or Turing tapes. For simplicity, we use the version which only has stacks. Although we can allow any number of stacks, \citet{book1970} show that we can convert any NLIN machine with $k > 3$ stacks into an equivalent machine with 3 stacks, or, equivalently 1 stack and 1 Turing tape (since two stacks can be used to simulate a Turing tape). We will therefore use, without loss of generality, the model with 3 stacks. The model with one stack corresponds to the standard stack automaton used to generate the context-free languages. While we know that machines like this with 2 stacks are already strictly stronger than the ones with only a single stack, it is unknown whether machines with 2 and 3 stacks are equivalent.

It is interesting to contrast NLIN to LOGCFL, a class more often found in mathematical linguistics. 


\begin{center}
\begin{tabular}{|l|l|l|l|}\hline
Class & stack size & Turing tape length & time \\ \hline
CFL & $n$ & none & polynomial \\
NLIN & $n$ & $n$ & linear \\
LOGCFL & $n$ & $log(n)$ & polynomial 	 \\ \hline
\end{tabular}
\end{center}

The main difference is that LOGCFG has a logarithmic space Turing tape and polynomial time for computing, whereas NLIN has a linear space Turing tape and linear time compute. 


An NLIN machine only makes a single left-to-right scan of its input, but it can store the input string on one of its stacks. Like a standard multi-tape Turing machine, it has a state and can scan the input symbol and the top symbols of all of the stacks.

\citet{book1970} prove a number of results about \nlin{} automata and the quasi-linear language generated by them: 
\begin{enumerate}
	\item we can convert any \nlin automaton into an automaton using only a constant number of steps between each scan of an input symbol \citep[Theorem~2.1]{book1970},
	\item we can convert any \nlin automaton $m$ into an equivalent automaton $m'$ such that $m$ accepts $w$ iff $m'$ accepts $w$ using exactly $2|w|$  steps \citep[Lemma~1.1]{book1970},
	\item  the quasi-linear languages $Q$ are the smallest AFL containing the context-free languages and closed under intersection (an AFL, an abstract family of languages, is a class of languages closed under union, concatenation, Kleene +,  non-erasing homomorphic mappings, inverse homomorphism and intersection with regular sets)  \citep[Corollary~1.2]{book1970},
\end{enumerate}

\subsection{First-order linear logic}

Figure~\ref{fig:fo} gives the cut-free sequent calculus rules for first-order linear logic. The reversible quantifier rules have the standard conditions that the eigenvariable $x$ of the eliminated quantifier does not occur free in $\Gamma$ or $C$. The irreversible quantifier rules $\forall L$ and $\exists R$ allow us to replace the quantifier variable $x$ by any term $t$. In practical theorem provers, we will generally use unification to find the most general term $t$ which makes the sequent provable during proof search. Even though complex terms are allowed in first-order linear logic, we only need constants in what follows.

\begin{figure}
\begin{center}
\begin{tabular}{ccc}
&\infer[\textit{Ax}]{A\vdash A}{} \\[2mm]
Irreverisble,  asynchronous && Reversible, synchronous \\[2mm]
\infer[\multimap L]{\Gamma, \Delta, B\multimap A\vdash C}{\Delta\vdash B & \Gamma,A \vdash C} &&
\infer[\multimap R]{\Gamma\vdash B\multimap A}{\Gamma,B\vdash A} \\[2mm]
\infer[\otimes R]{\Gamma,\Delta\vdash A\otimes B}{\Gamma\vdash A & \Delta\vdash B} &&
\infer[\otimes L]{\Gamma,A\otimes B \vdash C}{\Gamma,A,B \vdash C} \\[2mm]
\infer[\forall L]{\Gamma, \forall x. A\vdash C}{\Gamma, A[x:=t]\vdash C} &&
\infer[\forall R^*]{\Gamma\vdash \forall x. A}{\Gamma\vdash A} \\[2mm]
\infer[\exists R]{\Gamma\vdash \exists x.A}{\Gamma\vdash A[x:=t]} &&
\infer[\exists L^*]{\Gamma,\exists x.A\vdash C}{\Gamma,A\vdash C}
\end{tabular}
\end{center}
\caption{Sequent calculus rules for fist-order linear logic}
\label{fig:fo}
\end{figure}

As is clear from the rules, we use the intuitionistic, multiplicative fragment. Antecedents are multisets of formulas, so there is no need for an explicit commutativity rule. 

\subsection{Multitape online Turing machine}

An $n$-tape Turing machine is an $n+6$-tuple $\langle Q, \Sigma, n, \Gamma_1, \ldots, \Gamma_n, q_0,  F, \delta\rangle$ where:
\begin{description}
\item $Q$ is the finite set of states
\item $\Sigma$ is the finite set of input symbols
\item $n$ is the number of tapes
\item $\Gamma_i$ is the finite set of tape symbols for tape $1\leq i \leq  N$ 
\item $q_0 \in Q$ is that starting state
\item $F\subseteq Q$ is the set of final states
\item $\delta$ is the transition function.	
\end{description}


Let $\Sigma_{\epsilon}$ and $\Gamma_{\epsilon n}$ be, respectively the symbol sets $\Sigma \cup \{\epsilon\}$ and $\Gamma_i \cup \{\epsilon\}$ (for $1\leq i \leq  N$). Then the transition function $\delta$ is a function from $Q, \Sigma_{\epsilon}, \Gamma_{\epsilon1}, \ldots, \Gamma_{\epsilon n}$ to finite subsets of $Q, \Gamma_{\epsilon1}^*, \ldots, \Gamma_{\epsilon n}^*$ 

An instant description of such a machine is a tuple $\langle q_i, \Sigma^*, \Delta_1^*, \ldots, \Delta_n^*\rangle$, where the initial configuration is $\langle q_0, w, \epsilon, \ldots, \epsilon\rangle$ (for input string $w$ and $n$ empty stacks $\epsilon$) and final states are of the form $\langle q_F, \epsilon, \epsilon, \ldots, \epsilon\rangle$ (for $q_F \in F$, an no remaining symbols on the input string and the $n$ stacks).   

We interpret, given a state $q\in Q$, an input symbol $s\in \Sigma_{\epsilon}$,  
and the top symbols of the $n$ stacks,  it non-deterministically returns a new state and adds a fixed number of symbols to each stack. The symbol $\epsilon$ for the input means no symbol is read or removed. 


For a machine with $n$ stacks,  each transition changes a machine state as shown in~\ref{eq:trans} below.
\begin{equation}
\label{eq:trans}\langle q_i, as,  t_1b_1, \ldots, t_nb_n \rangle \rightarrow \langle q_j, s,  t_1u_1, \ldots, t_nu_n \rangle
\end{equation}

We look at the input symbol $a$ and the top stack symbols $b_i$ ($a$ and some or all of the $b_i$ can be $\epsilon$). Note that we use to convention that the topmost stack symbol is written at the rightmost position.

\begin{description}
\item $a \neq \epsilon$ Scan the input symbol and move right
\item $b_i \neq \epsilon$ Pop the top symbol from stack $i$
\item $u_i \neq \epsilon$ Push the sequence $u_i$ to stack $i$.
\end{description}

This is a standard definition of a multi-tape, online Turing machine where each tape is a stack. This restriction allows us to simplify the more general definition of  \citet{book1970}.

\subsection{Encoding NLIN machines and states into first-order linear logic}

For our encoding we use the following atomic formulas and basic predicates.

\begin{description}
\item  $\textit{word}(n, w, n+1)$ word $w$ occurs between positions $n$ and $n+1$ on the input string,\\
\item $\textit{head}(n)$ the read head is at position $n$ on the input tape,  \\
\item $\textit{stack}_i(n, s, n+1)$ symbol $s$ occurs on position $n$, $n+1$ of stack $i \in \{1,2,3\}$,  \\
\item $\textit{head}_i(n)$ the head of stack $i$ is on position $n$.
\item $q_i$, atomic formulas representing the current state $q_i$ of the Turing machine.
\end{description}

By convention, the empty input string corresponds to $\textit{word}(0,\epsilon,0)$,  and the empty stack $k$ to $\textit{stack}_k(0, \epsilon, 0)$



Our basic input is going to look like a standard first-order linear logic grammar,  with a sequence of input symbols $w_1,\ldots,w_n$, with each symbol $w_i$ assign position numbers $i-1,i$, from which we want to derive a  goal  goal state $\textit{state}(q_F)$.
\[
\textit{word}(0,w_0,1), \ldots,  \textit{word}(n-1,w_n,n)   \vdash \textit{state}(q_F)
\]

\subsection{Input sequence and goal  formula}

We will need to introduce a number of additional formulas,  representing the stacks and the Turing machine actions. We add all information for the starting state to the goal (succedent) formula.

For each symbol on the input sequence, we assign it a formula of the following form
\begin{align*} \textit{word}(i,s_i,i+1) \otimes A_1 \otimes \ldots \otimes A_k
\end{align*}
\noindent where $s_i$ is the symbol on the input tape between positions $i$ and $i+1$, and the $A_i$ are up to $k$ actions from the possible machine transitions.

This ensures we have a standard first-order linear logic lexicon with, by construction, a number of Turing machine operations which is linear on the size of the input string. When $n$ is the length of the input string and $k$ is the maximum number of operations added to lexical entries, the Turing machine performs at most $kn$ steps.

The goal formula adds the additional information needed to get the Turing machine into its initial state.  We need to do a number of things:
\begin{enumerate}
	\item enter the machine at its initial state
	\item set the head of the input tape to the first position of the input string
	\item initialise the three stacks to the empty stacks, with the stack head at the top.
	\item have a final state as goal
	\item require that the input has been fully read and that all stacks are empty at the end
\end{enumerate}



Empty stack $k$
\[\textit{head}_k(0) \otimes \textit{stack}_k(0,\epsilon,0) \]
Initial state (unique) 
\[q_0\]

Final state (non-deterministically chosen from the set of final states $Q_f$): 
\[q_F\]

Taking everything together, the goal formula will look as follows. 
\begin{align*}
q_0 &\multimap  \\
\textit{head}(0) \otimes \textit{word}(n,\epsilon,n) & \multimap \\
\textit{head}_1(0) \otimes \textit{stack}_1(0,\epsilon,0) &\multimap \\
\textit{head}_2(0) \otimes \textit{stack}_2(0,\epsilon,0) &\multimap  \\
\textit{head}_3(0) \otimes \textit{stack}_3(0,\epsilon,0) &\multimap \\
&q_F \otimes \\
& \textit{head}(n) \otimes \textit{word}(n,\epsilon,n) \otimes\\
&		 \textit{head}_1(0) \otimes \textit{stack}_1(0,\epsilon,0) \otimes \\
& \textit{head}_2(0) \otimes \textit{stack}_2(0,\epsilon,0) \otimes\\
& \textit{head}_3(0) \otimes \textit{stack}_3(0,\epsilon,0)
\end{align*}

The two $q_i$ atoms in the antecedent and the succedent ensure we
  start in the initial state $q_0$ and end in a final state $q_F$.
  
The $\textit{word}$ predicate in the antecedent adds $\epsilon$ at the end of the input string (position $n$).
The $\textit{head}(n) \otimes \textit{word}(n,\epsilon,n)$ subformula in the succedent ensures that the head is at the rightmost position at the end of the computation.

Finally, the  $ \textit{head}_k(0) \otimes \textit{stack}_k(0,\epsilon,0)$ formulas in both the antecedent and the succedent of the formula ensure that all stacks start and end empty.

In our representation, $\textit{head}_i(R)$, $R$ represents the rightmost (head) position of the stack. The initial position of each stack is $0$, other positions are unique variables generated by the quantifier rules.

\subsection{Turing machine operations}

Now that we have defined the basic structures and inputs, we define the machine operations as first-order linear logic formulas. The \emph{push} operation adds a number of symbols to one of the stacks, the \emph{pop} operation remove the top symbol from one of the stacks, and the \emph{scan} operation reads a symbol from the input string and moves right.

We start with the push operation, which is the most complicated one. First, we define an auxiliary \emph{push} predicate which we use to build to complete \emph{PUSH} formula.
Given a position $y$, we generate a sequence of $n>0$ symbols $s_1,\ldots s_n$ as follows 
		\begin{align*}
 \textit{push}_k(s_1\ldots s_n,y) = \exists x_1,\ldots x_n. &\textit{stack}_k(y,s_1,x_1) \otimes \ldots \otimes \\
&\textit{stack}_k(x_{n-1},s_n,x_n) \otimes \textit{head}_k(x_n)
\end{align*}
\noindent where 
$\textit{head}_k(x_n)$ will be the new stack head.



Push $s_1,\ldots,s_n$ on stack $k$ (this is the $b_k = \epsilon$ case).
\begin{align}
\textit{PUSH}_k(s_1\ldots s_n) =\forall y. \textit{head}_k(y) \multimap \textit{push}_k(s_1\ldots s_n,y)
\end{align}

We can define the $\textit{POP}_k(s)$ operation, which verifies the top symbol of stack $k$ is $s$ and remove it, as. follows.
\begin{align}
\textit{POP}_k(s) = \forall y z. \textit{head}_k(z) \multimap \textit{stack}_k(y,s,z) \multimap \textit{head}_k(y) 	  
\end{align}

Check the current input symbol is $s$ and move right.
\begin{align}
\textit{SCAN}(s) =\forall x y. \textit{head}(x) \multimap \textit{word}(x,s,y)  \multimap 
	\textit{head}(y)  
\end{align}

For convenience we can also add a formula \emph{REPL} which check the top symbol of stack $k$ is $s$,  then pops the top element and replace it by $s_1\ldots s_n$ (this is just a combination of $\textit{POP}$ and $\textit{PUSH}$).
\begin{align}
\textit{REPL}_k(s,s_1\ldots s_n) = \forall y z. \textit{head}_k(z) \multimap 	\textit{stack}_k(y,s,z) \otimes \textit{push}_k(s_1\ldots s_n,y)
\end{align}

The final type of formula we use is the state transition
\[
q_i \multimap q_j
\]

Given two formulas $F_1$ and $F_2$ of the form shown below. 
\[
F_1 = \forall x_0 \ldots x_n. (A_0 \multimap \ldots (A_m \multimap (B_0 \otimes \ldots\otimes B_k)))
\]
 and
 \[
 F_2 = \forall y_0 \ldots y_p. (C_0 \multimap \ldots (C_q \multimap (D_0 \otimes \ldots\otimes D_r)))
 \]
  We define their composition  $F_1 \circ F_2$ as follows.
 \begin{multline}\forall x_0 \ldots x_n y_0 \ldots y_p. (A_0 \multimap \ldots (A_m \multimap (C_0 \multimap \ldots (C_q \\
 \multimap (B_0 \otimes \ldots\otimes B_k) \otimes (D_0 \otimes \ldots\otimes D_r)))))
 \end{multline}
Now, each machine action is a combination of a state transition,
a possible input scan,  plus a possible stack operation for each  of the three stacks. 
We should be careful, though, since we cannot just combine the different elementary operations by `$\otimes$' since we need to enforce all different elementary operations which are combined into a single machine action are applied \emph{before} then next machine action. This is why we combine the operations using composition `$\circ$' instead.


Since each elementary machine action is translated by a formula of the form required for composition, we can combine these elementary actions into a single formula representing the transition.

Now let $A_1, \ldots,  A_k$ be the formulas corresponding to the $k$ transitions  in the Turing machine, and let $c$ be the constant such that the machine accepts all words $w$ in time $\leq c | w |$. The lexicon for our first-order linear logic grammar for each symbol will then be of the set of formulas $\textit{word}(i,s,j) \otimes B_1 \otimes \ldots \otimes B_d$ such that $d \leq c$, each $B_i $ is equal to $A_j$ for some $1\leq j\leq k$. In other words, when the character $s$ is part of the input string at position. $i-j$, then we select this word in combination with any number of machine actions up until the bound $c$. Note that it is possible to select the same action multiple times.

Similar to the linear-time context-sensitive languages, the resulting lexicon has a very high number of entries per character: given $k$ machine transitions and linear bound $c$, there will be $O(k^c)$ different lexical formulas for each character.

\begin{lemma} Let $m$ be a linear-time Turing machine with three stacks, let $l$ be the grammar in first-order linear logic corresponding to the translation of $m$, and let $s_1,\ldots,s_n$ be a word.  Then $m$ accepts $s_1,\ldots,s_n$ iff the corresponding string is derivable in the first-order linear logic grammar $l$.
\end{lemma}

\paragraph{Proof (sketch)} The lexicon and the goal formula generate the input configuration plus all linear choices of machine transitions.

The key invariant for showing we correctly implement machine actions like this is most easily shown by using focused proof search for a first-order linear logic. Given the way our translation creates formulas, different choices of focus correspond to different machine actions (among the linear number of action occurrences available at the start of the proof), and at the end of each synchronous phase and its subsequent asynchronous phase in focused proof search, exactly one machine transition has been performed.

Each selection of a focus formula corresponds to selecting which transition to apply. The positive $A$ subformulas will test whether the preconditions of the transition are satisfied. We defocus when we reach a negative $\exists$ or $\otimes$ formula. These are removed in the asynchronous phase, with $\exists$ rules creating new positions for material added to the stack,  and the $\otimes$ rules adding the information of the new machine state. \hfill $\Box$

 \paragraph{Example}
 The following automaton, with input state $q_0$ and output state $q_3$ generates the language $a^nb^nc^n$.
 \begin{align}
\label{trans:1} \langle q_0, \epsilon, \epsilon, \epsilon \rangle &\overset{\epsilon}{\rightarrow} \langle q_1, \epsilon, \epsilon, \epsilon \rangle \\
 \langle q_1, \epsilon, \epsilon, \epsilon \rangle &\overset{\epsilon}{\rightarrow} \langle q_2, \epsilon, \epsilon, \epsilon \rangle \\
 \langle q_2, \epsilon, \epsilon, \epsilon \rangle &\overset{\epsilon}{\rightarrow} \langle q_3, \epsilon, \epsilon, \epsilon \rangle \\
 \langle q_0, \epsilon, \epsilon, \epsilon \rangle &\overset{a}{\rightarrow} \langle q_0, d, d, \epsilon \rangle \\
 \langle q_1, d, \epsilon, \epsilon \rangle &\overset{b}{\rightarrow} \langle q_1, \epsilon, \epsilon, \epsilon \rangle \\
 \label{trans:end}\langle q_2, \epsilon, d, \epsilon \rangle &\overset{b}{\rightarrow} \langle q_2, \epsilon, \epsilon, \epsilon \rangle 
 \end{align}
 The operation is fairly simple: in its starting state $q_0$, it scans occurrences of $a$  and  pushed the symbol $d$ to the first two stacks for each $a$ encountered. 
 
 \begin{align}
 	\langle q_0, aabbcc, \epsilon, \epsilon, \epsilon \rangle  &\rightarrow \\ 
 	\langle q_0,abbcc, d, d, \epsilon \rangle &\rightarrow \\
    \langle q_0,bbcc, dd, dd, \epsilon \rangle &\rightarrow \\	
    \langle q_1,bbcc, dd, dd, \epsilon \rangle &\rightarrow \\	
    \langle q_1,bcc, d, dd, \epsilon \rangle &\rightarrow \\	
    \langle q_1,cc, \epsilon, dd, \epsilon \rangle &\rightarrow \\	    
   \langle q_2,cc, \epsilon, dd, \epsilon \rangle &\rightarrow \\	   
   \langle q_2,c, \epsilon, d, \epsilon \rangle &\rightarrow \\	   
    \langle q_2,\epsilon, \epsilon, \epsilon, \epsilon \rangle &\rightarrow \\	
     \langle q_3,\epsilon, \epsilon, \epsilon, \epsilon \rangle &\rightarrow
 \end{align}

The transitions in~\ref{trans:1} to~\ref{trans:end} are translated as linear logic formulas~\ref{ll:1} to~\ref{ll:end} below.
\begin{align}
 \label{ll:1} q_0 \multimap q_1 \\
 q_1 \multimap q_2 \\
 q_2 \multimap q_3 \\
\label{ll:scanpushpush} [q_0 \multimap q_0] \circ  \textit{SCAN}(a)\circ \textit{PUSH}_1(d) \circ \textit{PUSH}_2(d)  \\
 \label{ll:scanbpopd} [q_1 \multimap q_1] \circ   \textit{SCAN}(b)\circ \textit{POP}_1(d)\\
 \label{ll:end}       
   [q_2 \multimap q_2] \circ \textit{SCAN}(c)\circ \textit{POP}_2(d)
 \end{align}
Fully spelled out, formulas~\ref{ll:scanpushpush} to~\ref{ll:end}  look as~\ref{ll:scanpushpushfull} to~\ref{ll:scanendfull} below. 
\begin{multline}\label{ll:scanpushpushfull}
	\forall u v r y y'. q_0 \multimap \textit{head}(u) \multimap \textit{word}(u,a,v)  \multimap 	\textit{head}_1(y) \multimap \textit{head}_2(y') \multimap\\
	q_0  	\otimes \textit{head}(v) \otimes \exists z. \textit{stack}_1(y,d,z) \otimes \textit{head}_1(z) \otimes \\
	\exists z'. \textit{stack}_2(y',d,z') \otimes \textit{head}_2(z') 
\end{multline}
\begin{multline}
\label{ll:scanbpopdfull}	\forall u v y z.\ q_1 \multimap \textit{head}(u) \multimap \textit{word}(u, b, v)  \multimap 
\textit{head}_1(z)  \multimap \textit{stack}_1(y, d, z) \multimap \\
q_1 \otimes \textit{head}(v)  \otimes \textit{head}_1(y)\\
\end{multline}
\begin{multline}
\label{ll:scanendfull} \forall u v y z.\ q_1 \multimap \textit{head}(u) \multimap \textit{word}(u, c, v)  \multimap 
\textit{head}_2(z)  \multimap \textit{stack}_2(y, d, z) \multimap \\
q_1 \otimes \textit{head}(v)  \otimes \textit{head}_2(y)\\
\end{multline}

\subsection{Languages in NLIN}
\label{sec:langnlin}

We have already seen that we can generate $a^n b^n c^n$, one of the canonical mildly context-sensitive languages, in NLIN. So the natural question is whether we can do other canonical examples as well.

\paragraph{Counting dependencies}

First, note that we only needed two stacks for the language $a^n b^n c^n$, and that we can generalise this to more complex patterns of the form $a_1^n \ldots a_i^n$ for any $i$, by generalising the following strategy.

\begin{description}
	\item [Phase 1] Scan $a$, push $a$ on stack 1
	\item [Phase 2] Scan $b$, pop $a$ from stack 1, push $b$ to stack 2
	\item [Phase 3] Scan $c$, pop $b$ from stack 2, push $c$ to stack 1
	\item [Phase 4] Scan $d$, pop $c$ from stack 1
\end{description}

The basic idea is to change stacks at each change of symbol, pop the stack for the previous symbol to verify it occurs as many time as the current one, and push the current symbol onto the next stack to allow us to verify later it occurs as many times as the next symbol.

The fact that only two stack suffice to generate this entire hierarchy tells us that even when restricted to two stacks NLIN is already a very powerful tool for describing language classes.

\paragraph{Crossed dependencies}

Another $a^n b^m c^n d^m$

\begin{description}
	\item [Phase 1] Scan $a$, push $a$ on stack 1
	\item [Phase 2] Scan $b$, push $b$ on stack 2
	\item [Phase 3] Scan $c$, pop $a$ from stack 1
	\item [Phase 4] Scan $d$, pop $b$ from stack 2
\end{description}

In this case, it seems we need more than two stacks for the generalised versions of the language, like $a^n b^m c^n d^m e^n f^m$.
Adapting the previous strategy to this more complex example would require 4 stacks, but by \citet{book1970} we know we can reduce that number to 3 (for any number of stacks $> 3$).

\paragraph{Copy language}

\[
\{ ww \ |\   w \in (a|b)^* \}
\] 

\begin{description}
	\item [Phase 1] Scan $a$ or $b$, push the scanned symbol on stack 1
	\item [Phase 2] Scan $a$ or $b$, push the scanned symbol on stack 2
	\item [Phase 3] Until both stacks are empty,  pop the first symbol from both stacks and verify they are equal	
\end{description}

Generalising, we can use $n$ stacks for $n$ copies in the same way, and again use \citet{book1970} to limit ourselves to 3 stacks.

\paragraph{MIX}
 The automaton with single state $q_0$ and the following set of transitions generates the MIX language 
 \begin{align}
\label{mix:1} \langle q_0, a, b, c \rangle &\overset{\epsilon}{\rightarrow} \langle q_0, \epsilon, \epsilon, \epsilon \rangle \\
 \langle q_0, \epsilon, \epsilon, \epsilon \rangle &\overset{a}{\rightarrow} \langle q_0, a, \epsilon, \epsilon \rangle \\
 \langle q_0, \epsilon, \epsilon, \epsilon \rangle &\overset{b}{\rightarrow} \langle q_0, \epsilon, b, \epsilon \rangle \\
 \langle q_0, \epsilon, \epsilon, \epsilon \rangle &\overset{c}{\rightarrow} \langle q_0, \epsilon, \epsilon, c \rangle 
 \label{mix:end}
 \end{align}
It operates by pushing each of the symbols $a$, $b$, and $c$ to separate stacks, and allowing a pop operation only on all three stacks simultaneously.
\begin{align}
[q_0 \multimap q_0] \circ  \textit{POP}_1(a) \circ  \textit{POP}_2(b) \circ  \textit{POP}_3(c)	\\
[q_0 \multimap q_0] \circ  \textit{SCAN}(a) \circ  \textit{PUSH}_1(a) \\
[q_0 \multimap q_0] \circ  \textit{SCAN}(b) \circ  \textit{PUSH}_2(b) \\
[q_0 \multimap q_0] \circ  \textit{SCAN}(c) \circ  \textit{PUSH}_3(c) 
\end{align}

As before, we can use $n$ stacks for generalisations of MIX with $n$ symbols in the same way, then use \citet{book1970} to limit ourselves to 3 stacks.

\paragraph{Exponentials}

A standard non-semilinear language is the language $a^{2^n}$.

We can generate this language using two stacks and successively halving the input as follows.
\begin{description}
	\item Succeed if the input tape has a single symbol.
Scan the input tape, put every second symbol on stack 1
Fail if there is an odd number of $a$s.
\item Stack 1 now contains $n/2$ symbols.
\item Succeed if stack 1 has a single symbol, if not
Pop 2 symbols from stack 1, push 1 symbol on stack 2
Fail if stack 1 contains an odd number of symbols.
\item Continue alternating between stacks and halving until either a fail or a success state is reached 
\end{description}

The total time complexity of this is
\[
n + \frac{1}{2} n + \frac{1}{4} n + \frac{1}{8} n ...  = 2n
\]

\section{Discussion}

The quasi-linear languages of \nlin are a relatively powerful class of languages and it remains to be seen how \nlin relates to other complexity classes in formal language theory, like LOGCFL,  or to other language classes like RCL,  the class of languages generated by (positive) Range Concatenation Grammars, another powerful formalism proposed for natural language.

LOGCFL is the class of languages logspace reducible to the context-free languages. It is often considered a natural upper bound on the class of languages adequate for describing natural language.

Given that LOGCFL is a space complexity class and NLIN is a time complexity class, it would seem that proving a definite result relating the two classes would be very hard. It would similarly be hard to show that NLIN is a proper subset of the context-sensitive languages,  as this would require showing that non-deterministic linear time is included in linear space.

I believe that \nlin{} is an interesting alternative to classes like LOGCFL as an upper bound for formalisms intended to be applied to natural languages: the string languages generated by formalisms proposed for natural language should at most be in LOGCFL with respect to space complexity and at most in NLIN with respect to time complexity.

\subsection{Properties of formal languages}

\paragraph{Mildly context-sensitive languages} The mildly context-sensitive languages have a positive part and a negative part in their definition. 

The positive defining characteristic are
\begin{enumerate}
	\item generating the context-free languages,
	\item generating a limited set of non-context-free patterns such as $a^n b^n c^n$,  $a^n b^m c^n d^m$, and $\{ww | w \in \{a,b\}^*\}$.
\end{enumerate}

The negative defining characteristics are 
\begin{enumerate}
\setcounter{enumi}{2}
	\item\label{it:poly} polynomial fixed recognition, that is, when the grammar is fixed there is an algorithm in P which decides whether or not strings are in the language.
	\item  semilinearity, that is, the language generated is permutation-equivalent to a regular language. As a consequence, this means there are no `gaps' in the length between strings generated in the language $a^{n^2}$
\end{enumerate}
The formalisms we have seen in this paper, with the exception of the Lambek calculus, satisfy the positive characteristics but fail the negative ones yet still appear to be weaker than the context-sensitive languages. To make discussion easier, I will call formalisms which fail one of the two negative characterisations medium context-sensitive and those which fail both to be spicy context-sensitive. We will apply these adjectives only to classes which are subclasses of the context-sensitive languages with a fixed recognition problem in NP.




Given that polynomial fixed recognition generally comes with relatively high polynomials, typically at least $O(n^6)$, is is not clear to me that this is necessarily better than NP-complete for real-world applications.

I do agree that semilinearity is a desirable property to have and would love to find some restriction to type-logical grammars and first-order linear logic which guarantees it. This necessarily means abandoning intersection, but I believe the case in favour of intersection is much weaker than the case for semilinearity.

\paragraph{Homomorphism} One of the defining properties of being an AFL is closure under (non-erasing) homomorphisms. Homomorphisms have a surprisingly strong effect on the expressivity of formalisms like NLIN and RCG,  even when homomorphisms are restricted to non-erasing or even letter-to-letter homomorphisms. 

\cite{boullier00cubic} argues that a polynomial time parsing guarantee requires choosing between intersection and homomorphism and chooses intersection. For example, strings like $01\#p_1\vee p_2$ are logical formulas together with a certificate showing this formula to be decidable (in this case, setting $p_1$ to false and $p_2$ to true). Deciding whether a formula combined with an assignment to all its atomic formulas is satisfiable is something we can do in P and therefore with an RCG.  However, the letter-to-letter homomorphism $h(0) = ?$, $h(1) = ?$ transforms this string to $??\#p_1\vee p_2$. We have simply erased the assignments turning the strings of the polynomial language into a the classic NP-complete problem of satisfiability. Therefore, unless P=NP, RCG are not closed under homomorphisms.

For NLIN, we know that the intersections of context-free languages form an infinite hierarchy \citep{lw73intersect}, but that adding letter-to-letter homomorphism collapses this hierarchy at level 3 \citep[Theorem~4.1]{book1970}.

It is easy to see that lexicalised formalism like type-logical grammars are closed under letter-to-letter homomorphisms just by applying the homomorphism directly to the lexicon to produce the new grammar.

\paragraph{Permutation closure} In type-logical grammars, it has been known since the work of \cite{emmsextr} that adding extraction to the Lambek calculus gives us a simple way to add permutation closure to the generated language. The MIX language, the language consisting of equal numbers of $a$, $b$ and $c$ in any order, and its generalisations to $k > 3$ symbols are the standard example of languages which can be generated under permutation closure. We have seen in Section~\ref{sec:langnlin} that \nlin{} can generate $\textrm{MIX}_k$ for all $k$ and this is one of the reasons I believe this class is a good fit for first-order linear logic and possibly for other type-logical grammars as well.

In formal language theory, only very powerful formalisms like RCGs have an easy way to generate permutation closures. We know that tree adjoining grammars cannot generate MIX \citep{ks12mix}, although multiple context-free grammars can \citep{s15mix}. However, it is unclear whether these results generalise to arbitrary permutation closures. 

\paragraph{Semilinearity} Semilinearity is one of the defining properties of mildly context-sensitive formalisms. Semilinearity means the languages generated by a formalism are permutation equivalent to a regular language.  For example, the standard example of a mildly context-sensitive language which is not context-free is the language $a^n b^n c^n$ which is permutation equivalent to the regular language $(abc)^*$. By definition, semilinearity is invariant under permutation closure, so the MIX language and its generalisations are semilinear. MIX is permutation equivalent to the regular language $(abc)^*$ just like $a^n b^n c^n$.

\paragraph{Intersection} It is well-known that adding intersection to context-free grammars allows us to generate languages which are not semilinear.


For Range Concatenation Grammars, \citet{boullier00cubic} makes the choice to give up homomorphisms (and therefore being an AFL) to ensure polynomial parsing, noting that formalisms which have both have an NP-complete fixed recognition problems.

\nlin{} has homomorphisms and intersection and therefore NP-complete fixed recognition. Given that type-logical grammars trivially have letter-to-letter homomorphisms, the only solution available to avoid NP-complete fixed recognition in these formalisms is not adding intersection. This has the additional advantage of making semilinearity attainable again.

\newcommand{\newcrossmark}{\scalebox{0.85}[1]{$\times$}}
\begin{table}
\begin{center}
\begin{tabular}{|l|c|c|c|c|l|l|l|}\hline
& $h$ & $\cap$ & $\pi$ & semi & lang &	fr & ur\\ \hline
NLIN  & \checkmark & \checkmark & \checkmark & \newcrossmark & spicy & NP-complete & NP-complete \\
MCFG & \checkmark & \newcrossmark & ? & \checkmark & mild  & P & NP-hard \\
RCG & \newcrossmark &\checkmark & \checkmark & \newcrossmark & medium & P-complete & NP-hard \\
L  & \checkmark & \newcrossmark & \newcrossmark & \checkmark & context-free & P & NP-complete \\
MILL1  & \checkmark & \checkmark & \checkmark & \newcrossmark & spicy & NP-complete & NP-complete \\
TLG  & \checkmark & ? & \checkmark & \newcrossmark & spicy? & NP-complete & NP-hard \\ \hline
\end{tabular}
\end{center}
\caption{Formalisms discussed in this paper together with whether they are closed under homomorphisms $h$, intersection $\cap$, permutation $p$, whether the formalism only generates semilinear languages (semi) and the complexity of fixed and universal recognition (fr, ur)}
\end{table}

\subsection{Open questions}

\begin{enumerate}
\item Can we prove $L(\nlin) = L(MILL1)$ or $L(\nlin) \subsetneq L(MILL1)$?
\item What is the precise relation between NLIN and other formalisms like LOGCFL and RCL, or even the context-sensitive languages? Are there any languages generated by one of these formalisms but not the others? 
\item In formal language theory, can we separate permutation closure from other powerful operations such as copying and intersection? Since permutation closure does not contradict mild context-sensitivity, are there any mildly context-sensitive formalisms which are closed under permutation?
\item Can other type-logical grammars generate the quasi-linear languages? I think this might be possible for the Displacement calculus, using the separator symbols to mark the different stacks. 
\item Are there interesting fragments of type-logical grammars with polynomial fixed recognition but which still allow the standard type-logical treatment of linguistic phenomena like extraction and  quantification? I conjecture the answer to this question is no, mostly based on the fact that the known polynomial fragments are linguistically very restrictive.
	\item Are there interesting restrictions on first-order linear logic formulas and variables such that the resulting fragment is not closed under intersection and maybe even has polynomial fixed recognition? Like with other type-logical grammars, I am skeptical about polynomial fixed recognition, but I would like to see a fragment which is semilinear (and therefore not closed under intersection). This would necessarily mean \nlin{} is outside of this fragment.
	\item While letter-to-letter homomorphisms are trivial for type-logical grammars, do they actually form an AFL? I conjecture that they form a semi-AFL closed under concatenation but maybe not under Kleene plus as would be required for an AFL. Non-erasing homomorphisms are trivial. The Lambek calculus product makes inverse homomorphisms easy as well. Intersection with regular languages requires combining lexical entries $\textit{lex}(a) = F$ with left-linear rules $C \rightarrow aD$ to $(((C\backslash D)\backslash s) \Downarrow s)\bullet F$. Union and concatenation are simple as well using set union of the lexicon and disjoint sets of atomic formulas for the different grammars.
\end{enumerate}

\section{Conclusions}

In this paper, we have shown that type-logical grammars with the standard rules for product, associativity, extraction and infixation have an NP-complete fixed recognition problem, adapting a construction for first-order linear logic from \citet{pshenitsyn2025first}.

%


While NP-complete fixed recognition gives us information about the complexity class, it is not very informative about the class of formal languages generated by grammars in a formalism. \cite{pshenitsyn2025first} already shows that grammars in first-order linear logic are closed under intersection and  therefore not semilinear and not mildly context-sensitive. We have strengthened the results of \cite{pshenitsyn2025first} and shown that grammars in first-order linear logic can simulate the quasi-linear grammars of \citet{book1970}, the grammars recognised in non-deterministic linear time ($\textit{NSPACE}(n)$ or \nlin). I have argued that \nlin{} is an interesting class of formal languages as well, and a useful upper bound on the time complexity of formalisms used for natural language, just like LOGCFL is a useful upper bound on the space complexity.

 Taken together this means that first-order linear logic can be seen as the logic of non-deterministic linear-time string languages, graph languages and automata.



\bibliography{moot}
\bibliographystyle{agsm}
\end{document}